\documentstyle[11pt]{article} 
\input FEYNMAN
\title{\vspace{-1.in} \hfill {\small\rm TUM-HEP-304/97} \\~\\~\\
Dynamical Mass Generation in a \linebreak Finite-Temperature 
\linebreak Abelian Gauge Theory} 
\author{ George Triantaphyllou\thanks{e-mail:georg$@$physik.tu-muenchen.de}
$\;$\\{\it Alexander von Humboldt Fellow} \\~ 
\\{\it Institut f\"ur Theoretische Physik, Technische 
Universit\"at M\"unchen}\\
{\it James-Franck-Strasse, D-85748 Garching, GERMANY }}
\begin{document}
\setlength{\baselineskip}{23pt}
\maketitle
\begin{abstract}
We write down the gap equation for the fermion
self-energy in a finite-temperature 
abelian gauge theory in three dimensions. 
The instantaneous approximation is relaxed,
momentum-dependent fermion and photon 
self-energies are 
considered, and the corresponding Schwinger-Dyson
equation is solved numerically. The relation 
between the zero-momentum and zero-temperature fermion
self-energy and the critical 
temperature $T_{c}$, above which there is no dynamical 
mass generation, is then studied. We also 
investigate the effect which the number of fermion
flavours $N_{f}$ has on the results, and we give the 
phase diagram of the theory with respect to 
$T$ and $N_{f}$. 

~ 
\end{abstract}
\hspace{2.in} To appear in {\it Phys. Rev.} {\bf D}
\vspace{2.in}
\setcounter{page}{0}
\pagebreak

\section{INTRODUCTION}
In this paper we study dynamical fermion
mass generation in a three-dimensional abelian gauge theory
at finite temperature.   
The interest in phenomena associated
with finite-temperature strongly coupled systems
lies not only in their non-trivial field theoretical interpretation
but also in the fact that in some cases
specific comparisons between experimental
and theoretical results can be made. Dynamical fermion mass generation
is such a phenomenon whose role in such {\it a priori} unrelated effects
as chiral symmetry breaking in $QCD$ or superconductivity in solid
state systems gives it a particular importance. In the latter case,  
given the fact that certain copper oxides 
exhibit almost two-dimensional high-temperature superconductivity, the
study of $2+1$ field theories can be particularly instructive. 

Perturbation theory on its own
can unfortunately do little in exploring the
critical behavior of such systems. One has to use therefore
non-perturbative
techniques like effective potentials and Schwinger-Dyson (S-D)
equations, which,
although they cannot guarantee
the precision of their quantitative results,  
provide interesting qualitative insights. 

We proceed therefore by writing down the S-D equation for
the fermion self-energy in the real-time
formalism and then study its behaviour with temperature. 
In particular, of considerable interest is the quantity
$r = 2\Sigma(0)/k_{B}T_{c}$, where $\Sigma(0)$ is the fermion
self-energy at zero temperature and 
momentum and $T_{c}$ is the critical temperature  
above which there is no dynamical mass generation. The identification
of $\Delta = 2\Sigma(0)$ with the superconductor gap makes the model
in principle experimentally testable \cite{Mavro}. 

Similar investigations have been hitherto limited by the complexity
of the problem, in that the fermionic and photonic two-point functions
which appear in the gap equation depend not on only one variable
-as in the zero-temperature case- but on two independent ones. Responsible 
for that is the preferred reference frame (heat bath) associated with the
temperature, which breaks Lorenz invariance. Thus the quantities of
interest depend independently on energy and three-momentum.

Popular approaches so far use drastic approximations such as 
taking the fermion self-energy to be momentum independent and
truncating the energy dependence of the photon self-energies,
either by considering them as functions only of the four-momentum squared
\cite{ian}, 
or by taking the energy to be zero (the ``instantaneous approximation")
\cite{correct1}-\cite{correct2}.  

The study presented here is the first attempt to relax simultaneously
all these 
approximations by numerically solving the problem and keeping the
correct momentum dependence of the self-energies. This allows us to
explore the critical behaviour of the theory with respect to 
temperature and number of fermion flavours.
However, due to the complexity of the equations, we
neglect their imaginary parts, even though they
could play an important role \cite{pisar}. Moreover, we consider only
the one-fermion-loop contribution to the photon polarization diagrams,   
 an approximation justified for a large number of fermion flavours. 
 We also choose to truncate the
infinite S-D-equation hierarchy by replacing the full
photon-fermion vertex by the bare one. Being aware of the severity
of this approximation in relation to gauge-invariance and wave-function
renormalization \cite{penning}, 
we plan to make use of a more suitable truncation
of the S-D hierarchy in a future publication. 

\section{THE GAP EQUATION}

The Lagrangian of the three-dimensional
abelian gauge theory under study is a variation
of the usual $QED$ one with massless fermions:
\begin{equation}
{\cal L} = -\frac{1}{4e^{2}}F_{\mu \nu}F^{\mu \nu} + \bar{\psi_{a}}
(i\partial\!\!\!/- \tau_{3}A\!\!\!/)\psi_{a}
\end{equation}
\noindent where, as usual,  $F_{\mu \nu} = \partial_{\mu}A_{\nu} 
- \partial_{\nu}A_{\mu}$, $\alpha = 1,...,N_{f}$  with
$N_{f}$ the number of fermion flavours,  
the representation of the gamma matrices is four dimensional
(a possible choice is 
$\gamma_{0}= {\rm diag}(i\sigma_{3},-i\sigma_{3})$,
$\gamma_{1}= {\rm diag}(i\sigma_{1},-i\sigma_{1})$,
$\gamma_{2}= {\rm diag}(i\sigma_{2},-i\sigma_{2})$, where $\sigma_{i}$ are
the usual Pauli matrices)  
and
\[ \tau_{3} = \left( \begin{array}{cccc}
1&0& 0 & 0\\
0&1& 0 & 0\\
0&0&-1 & 0\\
0&0& 0 &-1
\end{array} \right) \]
Dynamical mass
generation in this case is parity conserving, which could be energetically 
preferred to a parity non-conserving one \cite{Appel1}.  
We will refer to the corresponding model as $\tau_{3}-QED$, as it
has been used in \cite{Mavro} in connection with high-temperature
superconductors.

The presence of the
$\tau_{3}$ matrix is responsible for the fact that 
this model does not have chiral symmetries that would be broken 
by a non-zero 
local order parameter like $<\bar{\psi}_{L}\psi_{R}>$, which
is associated with mass generation. 
Therefore, dynamical mass generation in this three-dimensional theory
can occur at finite non-zero temperatures without breaking 
any global symmetries
and therefore problems with the Mermin-Wagner
theorem \cite{mermin} are avoided. On the other hand, 
this will not influence the form of the
gap equation, which is the same as for usual $QED$.   

The corresponding S-D formalism for the two-point fermion function
in the rest frame of the heat bath
gives the following gap equation in the real-time formalism
for the fermion self-energy $\Sigma$,
for external momentum $p_{\mu} = (p_{0},\vec{p})$ and loop momentum
$k_{\mu} = (k_{0},\vec{k})$ in Euclidean space:
\begin{equation}
\Sigma(p_{0},|\vec{p}|)=\frac{ \alpha}{N_{f}}\int \frac{d^{3}k}{(2\pi)^{3}}  
D_{\beta}(k_{0},|\vec{k}|)S_{\beta}(p_{0} - k_{0},|\vec{p}-\vec{k}|) 
\label{eq:gap}
\end{equation}

\noindent 
where $D_{\beta}$ and $S_{\beta}$ are the photon and
fermion propagators, with the subscript $\beta$ indicating their
temperature dependence, and 
$\alpha = e^{2}N_{f}$ is the dimensionful coupling of
the superrenormalizable theory under study. 
 It can be considered as an
effective ultra-violet cut-off of the model \cite{Appel2}. 
In order to simplify the notation we write the first
argument of the various functions symbolically as  
$p_{0}$ instead of the more accurate $\sqrt{-p_{0}^{2}}$. 
Since we are
working at the one-loop level, an approach justified for large $N_{f}$,
we do not expect field-doubling
problems associated with the real-time formalism to play any direct role
\cite{doubling}. 

A bare photon-fermion vertex is used for simplicity, since
a more involved vertex would lead us to a system of coupled integral 
equations. The propagators $D_{\beta}$ 
and $S_{\beta}$ 
 in the Landau gauge at temperature $T$ are given by

\begin{eqnarray}
D_{\beta}(k_{0},|\vec{k}|) & = & 
\sum_{P=L,T} 
\frac{1}{k^{2}+\Pi_{P}(k_{0},|\vec{k}|)} 
+ \frac{2\pi \delta(k^{2}+\Pi_{P}(k_{0},|\vec{k}|))}{e^{\beta|k_{0}|} - 1} 
\nonumber \\
S_{\beta}(k_{0},|\vec{k}|) & = & 
\left(\frac{1}{k^{2}+ 
\Sigma^{2}(k_{0},|\vec{k}|)} 
- \frac{2\pi 
\delta(k^{2}+\Sigma^{2}(k_{0},|\vec{k}|)}
{e^{\beta |k_{0}|} + 1}\right) 
\Sigma(k_{0},|\vec{k}|)
\label{eq:prop}
\end{eqnarray}

\noindent
where $k^{2}=k_{\mu}k_{\mu}$, $\beta = 1/k_{B}T$, we sum over the
longitudinal and transverse photon polarizations $P=L, T$,
we have dropped the $k\!\!\!/$ term of the fermion propagator
since it disappears after the momentum integration, and we have
suppressed the imaginary $i$'s since they appear on both sides of the
gap equation.

The photon polarization functions $\Pi_{L,T}$ appearing above  
are given by \cite{ian} 
\begin{eqnarray}
\Pi_{L}(k_{0},|\vec{k}|) & = & \frac{\alpha k}{8} + \Pi_{1k} \nonumber \\
\Pi_{T}(k_{0},|\vec{k}|) & = & \frac{\alpha k}{8} - \Pi_{1k} + \Pi_{2k}, 
{\rm with} \nonumber \\ 
\Pi_{1k} & = & \frac{2\alpha k^{2}}{\pi \vec{k}^{2}}
\int \frac{d|\vec{p}|}{e^{\beta 
|\vec{p}|}+1}\left(1- \left(\frac{B_{k}+D_{k}}{2k^{2}}\right)^{1/2}\right)
\nonumber \\
\Pi_{2k} & = & 
\frac{2\alpha}{\pi} \int \frac{d|\vec{p}|}{e^{\beta|\vec{p}|}+1}
\left(1 - \left(\frac{k^{2}(B_{k}+D_{k})}{2D_{k}^{2}}\right)^{1/2}\right)
\nonumber \\
B_{k} & = & k^{2} - 4\vec{p}^{\;2} \nonumber \\
D_{k} & = & (B_{k}^{2} + 16k_{0}^{2}\vec{p}^{\;2})^{1/2} 
\end{eqnarray}

\noindent 
where the subscript $k$ is just a reminder
that the quantities are momentum dependent. It is worth noting that
these are calculated $via$ a one-loop massless fermion diagram, using
the fermion propagator of Eq.\ref{eq:prop}. 
The quantities $\Pi_{1,2K}$ are the finite-temperature contributions
to the photon polarization. They provide the thermal screening
responsible for the softening of the infrared behaviour of the theory,
since at small loop momenta they take values on the order  
of what is usually referred to as ``plasmon mass" squared 
$\omega^{2}_{P} =  \frac{\alpha \ln{2}}{\pi\beta}$ \cite{ian}.   
The one-loop approximation is justified for large $N_{f}$. However, the
masslessness of the fermions in this calculation
could introduce 
in principle consistency problems with Eq.\ref{eq:gap} which describes
fermion mass generation. We will return to this issue later. 

We are thus confronted with a three-dimensional non-linear integral equation
for a function of two variables. An analytical study of the full problem
seems an impossible task, so one is led to a computer simulation . Before
proceeding to a numerical solution however, we can try to somewhat
simplify the equation by noting that the delta function appearing in
the photon propagator gives a negligible contribution. In fact, this
function has two roots, one at very large momenta -of order $\alpha$-
where the fermion self-energy,
a decreasing function with momentum, is vanishingly small, and
one at very low momenta where the integrand is also small. Several authors
make the approximation of dropping this delta-function 
\cite{ian},\cite{delta1},\cite{delta2}, and we will also adopt it.  
The S-D equation therefore takes the following form
\begin{eqnarray}
\Sigma(p_{0},|\vec{p}|) &=& \frac{\alpha}{N_{f}}\int 
\frac{dk_{0}|\vec{k}|d|\vec{k}|d\theta}{(2\pi)^{3}}  
\frac{\Sigma(p_{0}-k_{0},|\vec{p}-\vec{k}|)}{(p-k)^{2}+ 
\Sigma^{2}(p_{0}-k_{0},|\vec{p}-\vec{k}|)} 
\times 
\sum_{P=L,T} 
\frac{1}{k^{2}+\Pi_{P}(k_{0},|\vec{k}|)} 
\nonumber \\ && \nonumber \\ 
 -  \frac{\alpha}{N_{f}} \int\frac{|\vec{k}|d|\vec{k}|d\theta}{(2\pi)^{2}}  
&& \!\!\!\!\!\!\!\!\!\!\!\frac{\Sigma(E,|\vec{p}-\vec{k}|)}
{2E(e^{\beta E} + 1)} 
 \times  \sum_{\epsilon=1,-1}\sum_{P=L,T} 
\frac{1}{(p_{0}-\epsilon E)^{2}+\vec{k}^{2}+
\Pi_{P}(p_{0}-\epsilon E,|\vec{k}|)} 
\label{eq:fingap}
\end{eqnarray}

\noindent where we sum over the photon polarizations $P=L, T$ and
over the two roots of the delta function by introducing $\epsilon=1,-1$, 
and $E$ is given by the relation 
$E^{2} = |\vec{p}-\vec{k}|^{2} + \Sigma^{2}(E,|\vec{p}-\vec{k}|)$.

We note that the definition of $E$ involves the fermion self-energy, which
we are trying to solve for, calculated at $p_{0}=E$. 
Therefore, we make the approximation 
$E \approx \sqrt{|\vec{p}-\vec{k}|^{2} + \Sigma^{2}(0,0)}$, which is valid
in the limits  
\footnote{We  assume of course that $\Sigma(0,|\vec{k}|) \approx
\Sigma(\Sigma(0,0),|\vec{k}|)$, since in the infrared, non-perturbative,
region our formalism does not have the required accuracy. 
We checked however that our final results do not
depend on this detail.}
$ |\vec{p}-\vec{k}|^{2} \gg {\rm or} \ll \Sigma^{2}(0,0)$. 
Another issue related to $E$ is that, for small
external momenta, we have to calculate the photon polarization functions
at regions which, if continued to Minkowski space, would make these
functions imaginary \cite{ian}. This is due to the fact that, whereas
we study fermion mass generation, the polarization functions were
calculated for massless fermions. In order to use consistently
massless fermions in connection with the photon polarization, we should 
really calculate the functions $\Pi_{P}$
at energies $p_{0} \pm |\vec{p}-\vec{k}|$ instead of  
$p_{0} \pm E$. We checked that both prescriptions give similar results. 

On physical grounds, we expect that increasing temperature ought to  
have decorrelating effects, which, after a certain point, should
make the order parameter $<\bar{\psi}_{L}\psi_{R}>$, and the
fermion self-energy associated with it, vanish.  
In fact, we see that 
the effect of finite temperature, both directly {\it via} the
negative second term of Eq.\ref{eq:fingap}, and indirectly {\it via} the
photon polarization functions is to reduce the integrand for a given
self-energy. It is therefore expected that, for large enough temperatures,
the self-energy will not find enough support from the right-hand side
and dynamical mass generation will be impossible, i.e. the only 
solution to the gap equation will be the trivial one. We proceed now
to verify -and quantify- this expectation. 

\section{NUMERICAL RESULTS} 
In order to attack the problem numerically, we have to
discretize our external and loop momentum space  
having in mind the characteristic energy scales of the model as well as
 the expected behaviour of the relevant quantities. The coupling 
 $\alpha$ sets an effective UV-cut-off, as has been seen in previous 
zero-temperature studies \cite{Appel2}. 
At small loop momenta, two quantities compete for
the role of a physical IR-cut-off: the fermion self-energy $\Sigma$
and $k_{B}T$, the energy
associated with the finite temperature. 
The latter would
become in effect only when both external and loop-momenta, as well
as the fermion self-energy, are small.
Problems related to the ambiguity of this cut-off due to the non-analyticities
of the photon polarization functions  are discussed in Ref.[2] as well as in 
Ref.\cite{nonan}. This  does not influence our results since in our study the 
fermion self-energy finally prevails as an IR-cut-off, as we will see
that convergence
of the algorithm is lost before $\Sigma$ becomes less important than $k_{B}T$. 
The gap equation is therefore both IR- and UV- finite, and 
we will use IR and UV cut-offs only for numerical reasons. 
Noting that $\Sigma$ is a function
decreasing with momentum, and that the expected hierarchy of scales is 
large (three orders of magnitude or more), we discretize
$p^{2}_{0}$, $|\vec{p}|^{2}$, $k^{2}_{0}$, $|\vec{k}|^{2}$ according to
$\log_{10}{(\Lambda^{2}_{IR})} + \frac{i}{n}\log_{10}{(\Lambda^{2}_{UV}/
\Lambda^{2}_{IR})}$ and the angle $\theta$ according to
$2\pi i/n$, where $i = 1, ..., n$. 
We do not take the points $\theta = 0, \pi$
as integration points, since the  kernel has an integrable singularity
there. 

We thus have a lattice with five dimensions, two coming
from the external and three from the loop momenta. 
Furthermore, we are always careful that 
$\Lambda_{IR} < k_{B}T, \Sigma(0,0)$ (by $\Sigma(0,0)$ of course we 
mean here and in the following
$\Sigma(\Lambda_{IR},\Lambda_{IR})$),
and $\Lambda_{UV}> \alpha$, and check the dependence of our results
on the particular values of the cut-offs and the size of the momentum
lattice, which we took to be $8^{5}$, $12^{5}$
and  $16^{5}$. Even though 
the quantities $\Sigma(0,0)/\alpha$ 
and $k_{B}T_{c}/\alpha$ vary with different
choices of cut-offs or relaxation-speed parameter, 
the ratio $r=2\Sigma(0,0)/k_{B}T_{c}$, remains pretty stable. 

The numerical relaxation method employed to solve the
gap equation consists of inserting an initial
``input" configuration for the fermion self-energy $\Sigma$ 
to (\ref{eq:fingap}), taking the ``output" configuration as the
new ``input", and then iterating the equation until it is 
satisfied to a good accuracy. In particular, we consider that the
algorithm has converged when the mean difference as well as
the standard deviation of the points of the ``input"
and ``output" functions is less than $10\%$. This is about
the best accuracy our algorithm can achieve at zero temperature
and $N_{f} = 1$.
It should be noted that in all cases studied, for
sufficiently small values for the temperature and 
fermion flavors, the mean difference of the
input and output configurations converges much faster to zero 
than the standard deviation between them. This allows us to distinguish  
the deviation caused by numerical error from the deviation caused by
a possible overall fall of the solution towards zero. One can then be
confident that the solution found is stable and not converging
to the trivial one. 
In order to avoid 
convergence to unstable, oscillating solutions corresponding to
small $\Sigma(0,0)$ \cite{Appel2},
we take the original configuration to be flat and equal to $\alpha$. 
Initial configurations falling smoothly with momentum only
decrease the convergence time and are therefore preferable, 
since they are closer to the final
solution, and they give the same results.

The fermion and photon self-energies are needed inside the integrand on
points outside the original momentum lattice. Their corresponding values
are found by linear interpolation from the values of the quantities on the 
lattice. For large internal and external momenta, some arguments can
fall also outside the lattice space. Even though we use some extrapolation
to recover the 
functions with such arguments, the values of the self-energies
for momenta near the UV-cut-off, even though small
and insignificant for the final results, should be taken with caution. 
Moreover, in order to obtain a smooth
solution we take the values of the output function at an 
external momentum lattice point $(i,j)$ to be the average of
the integration result $\Sigma_{{\rm out}}$, 
i.e. 
$\Sigma(i,j)=\left(
\Sigma_{{\rm out}}(i+1,j)+\Sigma_{{\rm out}}(i-1,j)\right.$
$\left.+\Sigma_{{\rm out}}(i,j+1)+\Sigma_{{\rm out}}(i,j-1) \right)/4.$
This is just one of several similar stabilizing procedures one
could use, and we have checked 
that the results do not depend on the particular
choice of such a procedure. 

\begin{figure}[t]
\vspace{4.5in}
\includegraphics{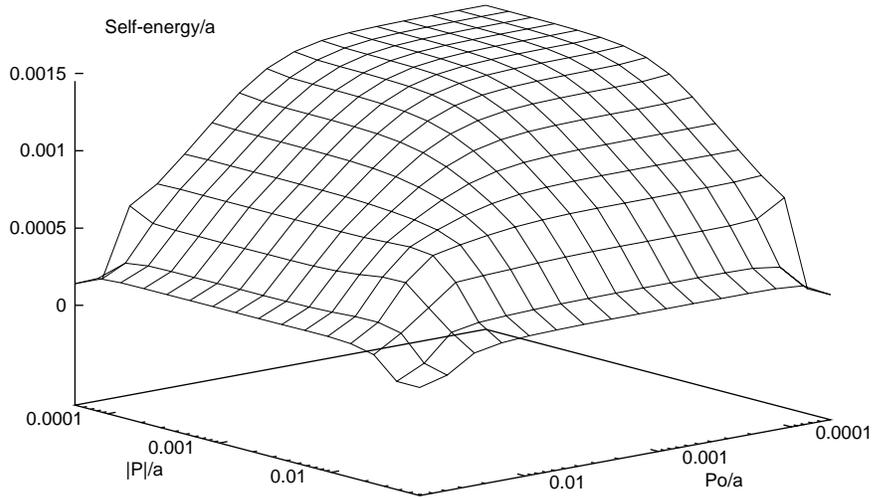}
\caption{The self-energy $\Sigma(p_{0},|\vec{p}|)$ at $T=0$ and
$N_{f}=2$ for a $16^{5}$ lattice, plotted
as a function of momenta $p_{0}$ and $|\vec{p}|$ in logarithmic scale.
All quantities are scaled by the coupling $\alpha$. 
}
~\\
\label{fig:rhofig1}
\end{figure}
For $T = 0$ and $N_{f} = 2$ we get the solution shown in Fig. 1.
It exhibits roughly the expected behaviour $\Sigma(p) \approx 
\frac{\sin{\left[\gamma\{\ln{(p/\Sigma(0))}+ \delta\} \right]}}{p^{1/2}}$
for
$\Sigma(0) < p < \alpha$,  
 with $\gamma$ and $\delta$ in principle
functions of $N_{f}$ \cite{Appel2}. We find that the general form
of $\Sigma(p_{0},|\vec{p}|)$ does not change 
with various $T$ or $N_{f}$. 

When the temperature exceeds some critical value $T_{c}$, the
algorithm does not converge, i.e. the deviation of the ``output" from
the ``input" function oscillates enormously from iteration to iteration,
both in magnitude and in sign, 
instead of decreasing monotonically and reaching the required $10\%$ value.
For these high temperatures,
 even though the algorithm does not 
converge, the ``output" configuration
for the self-energy after each iteration is tending on the average
 rapidly to  zero (the higher the temperature, the more rapid the fall). 
Stating this differently, only a trivial input configuration seems to
be able to satisfy the self-consistency of the gap equation. 
On the other hand, for temperatures
just below critical, $\Sigma(0,0)$ is still at about the same the
order of magnitude as its zero-temperature value.  
This does not necessarily 
mean that the self-energy drops ``immediately to zero"
when the critical temperature is reached. It most probably means
that we are slightly underestimating $T_{c}$, since
at temperatures just below the real critical value  it is very difficult to 
find an input configuration irregular enough to achieve
convergence to the solution, which in the neighbourhood of $T_{c}$
is probably oscillating and on the average much smaller in magnitude.  
   \begin{table}[t]
    \begin{tabular}{||c
 ||@{\hspace{2mm}}c@{\hspace{2mm}}|@{\hspace{2mm}}c@{\hspace{2mm}}
 |@{\hspace{2mm}}c@{\hspace{2mm}}|@{\hspace{2mm}}c@{\hspace{2mm}}
 |@{\hspace{2mm}}c@{\hspace{2mm}}|@{\hspace{2mm}}c@{\hspace{2mm}}
 |@{\hspace{2mm}}c@{\hspace{2mm}}|@{\hspace{2mm}}c@{\hspace{2mm}}
 |@{\hspace{2mm}}c@{\hspace{2mm}}||}  \hline
\rule[-3mm]{0cm}{8mm}
Fermion flavours $\Rightarrow$
&\multicolumn{3}{c||@{\hspace{2mm}}}{$N_{f}=1$} &  
 \multicolumn{3}{c||@{\hspace{2mm}}}{$N_{f}=2$} & 
 \multicolumn{3}{c||}{$N_{f}=3$}  
  \\  \cline{1-10} 
\rule[-3mm]{0cm}{8mm}
Lattice size $\Downarrow$ 
&$s_{o}$&$t_{c}$&$r$&$s_{0}$&$t_{c}$&$r$&$s_{0}$&$t_{c}$&$r$ \\ \cline{1-10}
\rule[-3mm]{0cm}{8mm}$8^{5}$ &15&3.8&7.9&3.1&0.6&10.3&0.9&0.17&10.6  
\\ \cline{1-10}
\rule[-3mm]{0cm}{8mm}$12^{5}$ &19&4.5&8.4&3.8&0.67&11.3&0.68&0.11&12.4  
\\ \cline{1-10}
\rule[-3mm]{0cm}{8mm}$16^{5}$&21&4.8&8.7&2.4&0.4&12&0.23&0.035&13.1 
\\ \hline \hline 
      \end{tabular}  
 \caption{ The quantities $s_{0}=10^{3} \times \Sigma(0,0)/\alpha$
 at $T=0$, $t_{c}= 10^{3} \times k_{B}T_{c}/\alpha$ and 
 $r=2s_{0}/t_{c}$ for different numbers of fermion 
flavours and lattice sizes $n^{5}$ and for $\Lambda_{UV}/\alpha = 0.1$. } 
~\\
\label{table:rhota}
     \end{table}

   \begin{table}[t]
    \begin{tabular}{||c
 ||@{\hspace{2mm}}c@{\hspace{2mm}}|@{\hspace{2mm}}c@{\hspace{2mm}}
 |@{\hspace{2mm}}c@{\hspace{2mm}}|@{\hspace{2mm}}c@{\hspace{2mm}}
 |@{\hspace{2mm}}c@{\hspace{2mm}}|@{\hspace{2mm}}c@{\hspace{2mm}}
 |@{\hspace{2mm}}c@{\hspace{2mm}}|@{\hspace{2mm}}c@{\hspace{2mm}}
 |@{\hspace{2mm}}c@{\hspace{2mm}}||}  \hline
\rule[-3mm]{0cm}{8mm}
Fermion flavours $\Rightarrow$
&\multicolumn{3}{c||@{\hspace{2mm}}}{$N_{f}=1$} &  
 \multicolumn{3}{c||@{\hspace{2mm}}}{$N_{f}=2$} & 
 \multicolumn{3}{c||}{$N_{f}=3$}  
  \\  \cline{1-10} 
\rule[-3mm]{0cm}{8mm}
 $ \Lambda_{UV}/\alpha \; \Downarrow$ 
&$s_{o}$&$t_{c}$&$r$&$s_{0}$&$t_{c}$&$r$&$s_{0}$&$t_{c}$&$r$ \\ \cline{1-10}
\rule[-3mm]{0cm}{8mm}$10^{-2}$ &9.3&2.3&8.1&0.9&0.17&10.6&0.1&0.016&12.5  
\\ \cline{1-10}
\rule[-3mm]{0cm}{8mm}$10^{-1}$ &21&4.8&8.7&2.4&0.4&12&0.23&0.035&13.1  
\\ \cline{1-10}
\rule[-3mm]{0cm}{8mm}1&23&5.1&9&3.3&0.54&12.2&0.5&0.074&13.5 
\\ \hline \hline 
      \end{tabular}  
 \caption{ The quantities $s_{0}=10^{3} \times \Sigma(0,0)/\alpha$
 at $T=0$, $t_{c}= 10^{3} \times k_{B}T_{c}/\alpha$ and 
 $r=2s_{0}/t_{c}$ for different values of  
flavours and ultra-violet cut-offs $\Lambda_{UV}$, for a $16^{5}$ lattice.}  
~\\
\label{table:rhota2}
     \end{table}

We list in Table 1 the values
of the fermion self-energy $\Sigma(0,0)$ at zero momentum and temperature,
of the critical temperature, both scaled by $\alpha$, and of the ratio $r$,
for different values of the number of fermions $N_{f}$ and lattice sizes.
We choose the UV-cut-off such that $\Lambda_{UV}/\alpha = 0.1$, so that
not only it is smaller than the physical cut-off $\alpha$, but also 
truncates the integration  before  the uninteresting region where the
self-energy is vanishingly small.
Even though the $\Sigma$ and $T_{c}$ values do not follow a specific
pattern with respect to the lattice size $n^{5}$, 
the $r$ ratio seems to be converging for larger $n$. 
In particular, the difference between the $r$-values for
the $12^{5}$ and $16^{5}$ lattice is consistent with the $10\%$
accuracy provided by the numerical algorithm.  However, given the large
scale hierarchies of the problem, the $n=8$ case for instance
is just indicative. 
Moreover, since we justified the one-loop 
calculations with the $1/N_{f}$ expansion,
the small values of $N_{f}$, and especially the ones
 reported for $N_{f} = 1$, should be taken with caution.
The values for $N_{f}=2$ are the ones relevant to the high-$T_{c}$
superconductors.
Furthermore, we see that for $N_{f}=3$ the hierarchy between 
$\Sigma(0,0)$ and $\alpha$ is very large. 
The relatively small lattice size forced upon us by computer-power limitations 
might therefore explain the stronger dependence
of $s_{0}$ and $t_{c}$ on the lattice size for three fermion flavors. 

In Table 2 we present our results for a $16^{5}$ lattice for 
different  ratios $\Lambda_{UV}/\alpha$. The ratio
$r$ follows a similar pattern for various UV-cut-offs, 
which is expected from a super-renormalizable theory. Lower values
for $\Lambda_{UV}$ would truncate the integration too early, since the
self-energy would be non-negligible there. When $\Lambda_{UV}$ is
taken to be larger than $\alpha$, the algorithm does not converge, showing 
that it is not easy to perform the integrations
on momenta above the physical cut-off $\alpha$ with
such small lattices. The reason is that
numerical errors dominate the calculation at this region, since  
the self-energy, a function decreasing with increasing momenta,
is very small there (typically $\Sigma(0,0)/\alpha$
is on the order of $10^{-2}$ and smaller). 

We see that the ratio $r$ increases with increasing $N_{f}$.
This behavior, as well as the overall magnitude of $r$,
is consistent with the results of Ref.\cite{ian} for the case
including retardation, when 
the photon polarization functions are approximated by 
$\Pi_{L} \approx \Pi_{T} \approx \frac{\alpha k}{8} +2\omega_{P}^{2}
\;\;{\rm or}\;\; \frac{\alpha k}{8} +\omega_{P}^{2}$, 
as in \cite{delta1}. Since in our treatment the exact one-loop photon
polarizations for massless fermions are used, our results indicate that
this is the most sensible approximation. The other approximations 
in \cite{ian} get smaller r ratios by increasing at the same time
the value of the critical
temperature by a factor as large as
2 or 3, which renders their validity questionable.
 An indication that
the results in Tables 1 and 2 cannot be easily compared to the ones in 
\cite{ian} for various numbers of fermion 
flavors is the fact that in that paper for instance  
results for $N_{f}=4, 5$ are reported, whereas in our case the
 behaviour of the theory with $N_{f}$  renders  
mass generation with
so many fermion flavors difficult or very small, which is in accordance with
theoretical expectations. A more accurate study of the behaviour
of the theory with $N_{f}$ is enabled by the relaxing of several 
approximations used in \cite{ian}, and is presented later in this paper.

The values given for $r$ could be overestimating this ratio, as
a result of several effects. First, since $\Sigma$ is momentum dependent, 
one could take 
the relevant value for the ratio $r$ to be $\Sigma(p=\Sigma(0))$ instead
of $\Sigma(0)$. This could decrease $r$ by $10-20\%$, which is justified
not only by the accuracy of our algorithm but also by our limited
understanding of infrared dynamics. Second, as was already noted, 
at the temperature 
at which fluctuations destroy the convergence of our algorithm,
$\Sigma(0,0)$ has still the same order of magnitude as its
value at zero-temperature. This not
only shows that the fermion self-energy falls pretty abruptly when
the temperature reaches its critical value, but that we might be
slightly underestimating $T_{c}$. 
The self-energy might be non-zero and 
falling with temperature for somewhat higher
temperature values by taking highly irregular shapes
that our algorithm is unable to find.
In addition, it might not be fair to
expect the algorithm at zero temperature and at temperatures close to the
critical value to reach the same accuracy of $10\%$, which after all 
determines which value of critical temperature we report, since
larger temperatures make the numerical error coming from the
second term of the right-hand side of Eq.\ref{eq:fingap} more
important. 
\begin{figure}[ht]
\vspace{4.5in}
\includegraphics{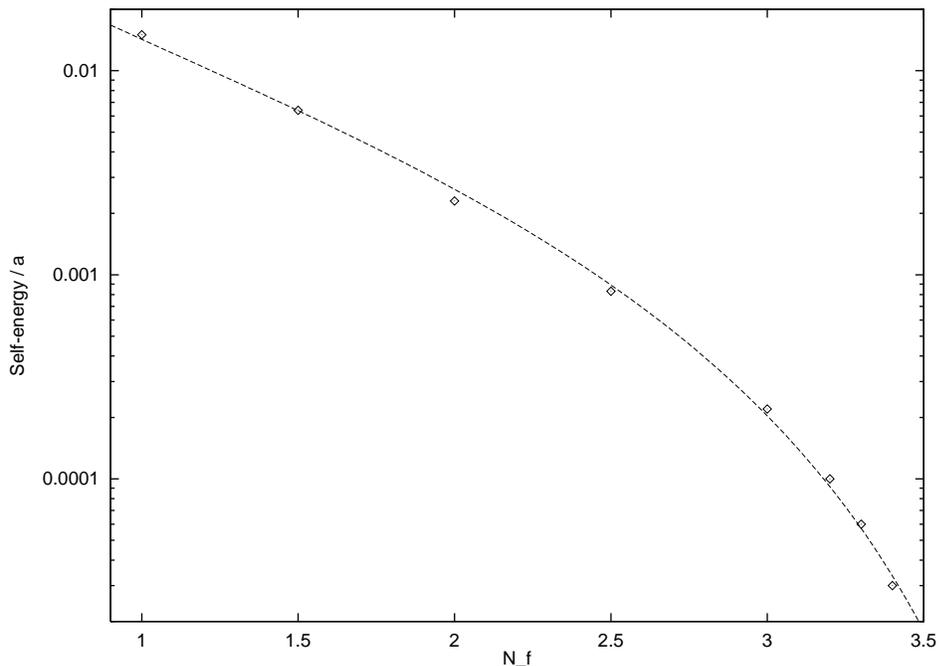}
\caption{The fermion self-energy at zero momentum and zero temperature, 
scaled by $\alpha$ and on a logarithmic scale,  
with respect to
$N_{f}$ for a $16^{5}$ lattice. 
We fit our results with the curve 
$\Sigma(0,0)/\alpha = \exp{\{-4.5/(N_{c}/N_{f}-1)^{1/2}\}}/6$, 
with $N_{c}=4.35$. Values of $N_{f}$ larger than 3.4 are not
considered, because then the self-energy falls below the IR-cut-off.}  
~\\
\label{fig:rhofig2}
\end{figure}

We found that for large values of $N_{f}$ the algorithm is also not 
converging. 
As in the critical temperature case, failure of convergence is 
accompanied with a rapid decrease of the output configurations, which
is a sign that the solution tends to the trivial one.
This indicates that there might be a critical value $N_{c}$ 
above which dynamical mass generation is impossible. 
In particular, for a lattice size of $16^{5}$ we found that
$N_{c} = 3.4$ for $k_{B}T \approx \Sigma(0,0) \approx \Lambda_{IR}
\approx 10^{-5}\times \alpha$.
Larger values of $N_{f}$ not only produce fermion self-energies 
tending below the IR-cut-off, but are also unable to reach the
required accuracy of $10\%$. This value is consistent
with theoretical expectations which give $N_{c} = 32/\pi^{2}$
at zero temperature \cite{Appel2}. 

In Fig. 2 we plot the fermion
self-energy at zero momentum and zero temperature
as a function of $N_{f}$, and we fit it
with the -phenomenological- curve 
$\Sigma(0,0)/\alpha = \exp{\{-4.5/(N_{c}/N_{f}-1)^{1/2}\}}/6$. 
From the fit a value $N_{c}=4.35$
seems to be close to our data. The non-analytic form of the fitted
curve is needed not only to describe the exponential fall-off
of the self-energy but also its probable
vanishing at the critical value of $N_{f}$,
and it exhibits the non-perturbative character of the 
dynamics. However, the approach of the self-energy to $\Lambda_{IR}$
does not allow us to draw firm conclusions on the exact 
vanishing of $\Sigma$ at a critical number of fermion flavors. 
The deviation of this formula from theoretical expectations
which give it as 
$\Sigma(0,0)/\alpha \approx \exp{\{-2\pi/(N_{c}/N_{f}-1)^{1/2}\}}$ 
with $N_{c} = 32/\pi^{2}$ \cite{Appel2} is most probably 
due to the fact that the latter non-analytic formula is intended only for
$N_{f}$ very close to $N_{c}$, and is not expected to fit data with
smaller $N_{f}$.
\begin{figure}[ht]
\vspace{4.5in}
\includegraphics{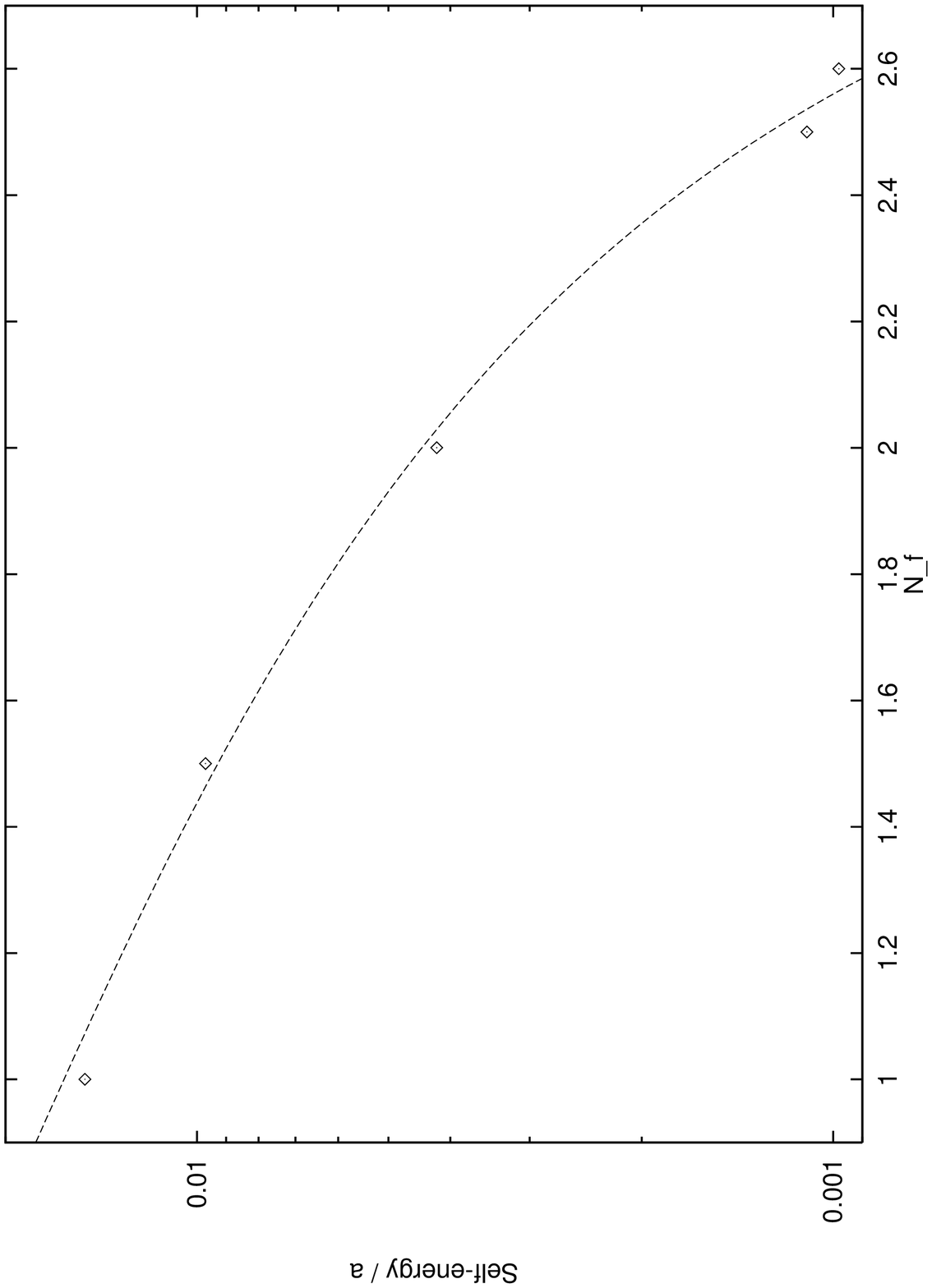}
\caption{The self-energy at zero momentum and at $k_{B}T/\alpha= 10^{-4}$,
scaled by $\alpha$ and on a logarithmic scale,
with respect to
$N_{f}$ for a $16^{5}$ lattice. 
We fit our results with the curve 
$\Sigma(0,0)/\alpha = \exp{\{-2.1/(N_{c}/N_{f}-1)^{1/2}\}}/15$, 
with $N_{c}=3.2$. The algorithm does not converge for values of 
$N_{f}$ larger than 2.6.}  
~\\
\label{fig:rhofig3}
\end{figure}

In Fig. 3 we proceed in a similar manner for a non-zero temperature
case, namely $k_{B}T_{c}/\alpha = 10^{-4}$. We fit the zero-momentum
fermion self-energy with the -phenomenological- curve
$\Sigma(0,0)/\alpha = \exp{\{-2.1/(N_{c}/N_{f}-1)^{1/2}\}}/15$. 
From the fit a value $N_{c}=3.2$ is favored. However, we loose 
convergence of our algorithm
at $N_{f} \approx 2.6$, which should be much closer
to the real critical value $N_{c}$, according to the criteria
used so far. This deviation indicates, as
in the zero-temperature case, that the non-analytic 
functional form of the fit
should not be used for values of $N_{f}$ far away from $N_{c}$. In this
finite-temperature case, convergence of the numerical algorithm
is lost before the self-energy falls below the IR-cut-off.
Figures 2 and  3 are similar to the corresponding ones in \cite{correct2}, 
which also indicate the existence of a critical behaviour with $N_{f}$,
even though our value for $N_{c}$ at  zero temperature is closer to
the theoretical expectations. 
The decrease of $N_{c}$ with increasing temperature is also  
expected,  and is shown in more detail in the following 
figure, i.e. the phase diagram.  
\begin{figure}[ht]
\vspace{4.5in}
\includegraphics{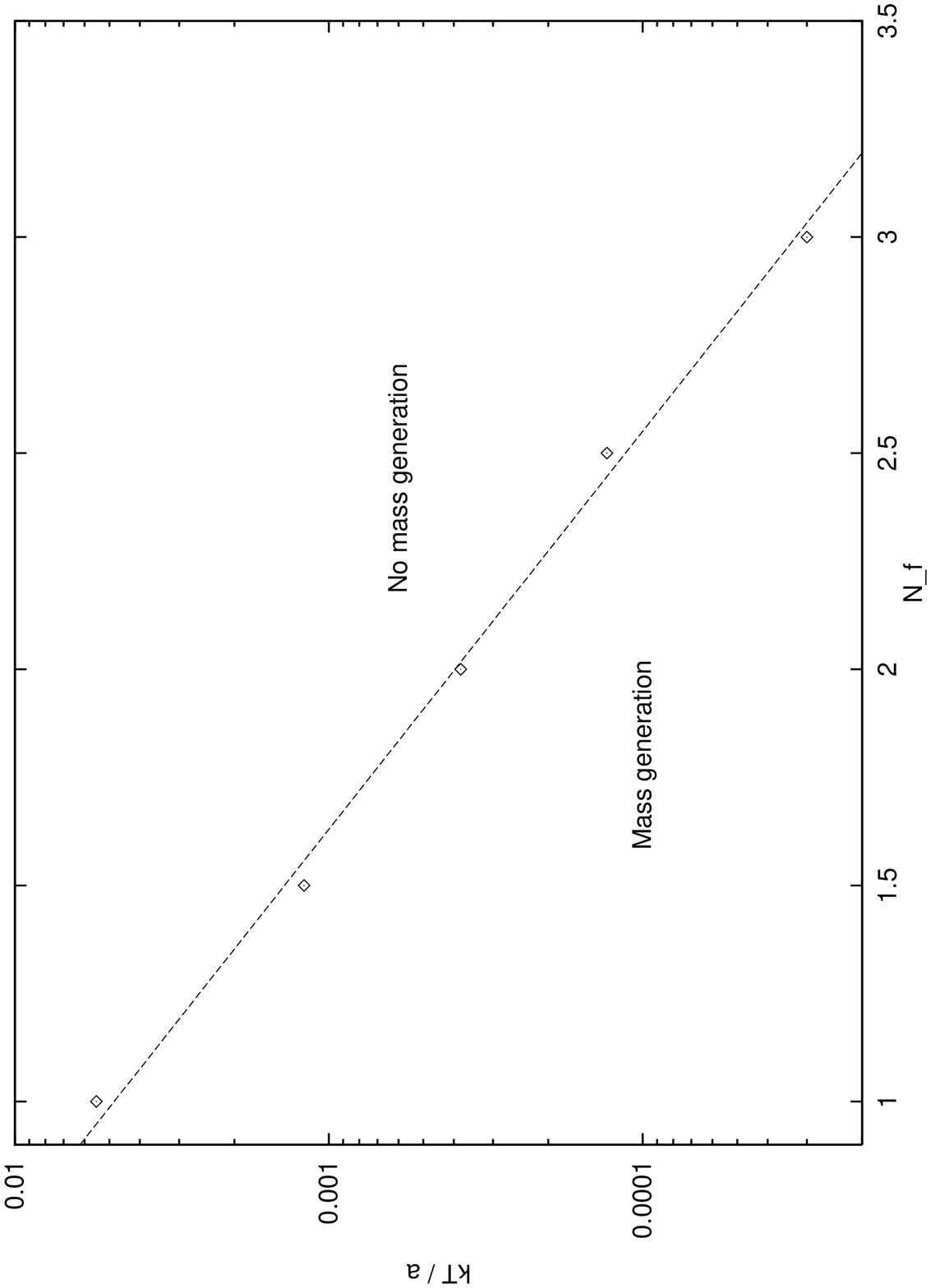}
\caption{The phase diagram of the theory with respect to
$k_{B}T$ on a logarithmic scale and $N_{f}$ for a $16^{5}$ lattice. 
We fit the critical line with the curve
$k_{B}T/\alpha = \exp{\{-2.5 N_{f}\}}/17$. This functional form
could lose its validity  for $N_{f}$ larger than about 3, since
then $T_{c}$ falls below the IR-cut-off. 
}
~\\
\label{fig:rhofig4}
\end{figure}

In Fig. 4 we plot the phase diagram of the theory 
with respect to $N_{f}$ and $k_{B}T$, where by the term  
``phase" we mean a situation where there is or there is
no dynamical mass generation.
The non-perturbative nature of our formalism allows us to consider also
non-integer  values of $N_{f}$. We were able to fit the critical line with
the -phenomenological- curve $k_{B}T_{c}/\alpha = \exp{\{-2.5N_{f}\}}/17$.
A similar exponential fall-off has already been seen in \cite{correct2}.  
This functional form should not be used for values of $N_{f}$ larger
than about 3, because then $T_{c}$ falls below the IR-cut-off.
It is quite possible that when $N_{f} \approx 3.4$ the behavior of $k_{B}T$
with $N_{f}$ becomes here also non-analytic. 
Nevertheless, we are not able to explore this region, since
for values of $N_{f}$ larger than about 3,  
$kT_{c}$ falls below the IR-cut-off. 

\section{CONCLUSIONS}
We report results indicating that for temperatures below a finite
critical value, dynamical mass generation in finite
temperature $\tau_{3}-QED$ in three dimensions is possible. 
We find a large ratio $r$ that not only exceeds 
the usual BCS prediction, but also the value measured in 
high-temperature superconductors, which is close to 8 \cite{superc}, and
indicates anyway that these are very strongly coupled systems. 
This ratio, on the order of 10, is generally consistent with results of 
previous studies \cite{ian}, even though its precise behaviour with
$N_{f}$ shows  some differences. 
We also find that for a number of fermion flavours larger than a critical
value, of order 3,
there cannot be mass generation, or, even if there is any, it
falls below any IR-cut-off that we may set so it is impossible to study 
with the present algorithm. We are then able to draw the phase
diagram of the theory, with respect to temperature and fermion
flavours, which separates the regions where there is and where 
there is not dynamical mass generation.

We were able to go beyond the 
instantaneous approximation and
consider momentum dependent fermion and photon self-energies 
by working with quantities depending on two independent variables 
simultaneously. 
Previous results on this context,
relying on severe truncations and approximations, 
are thus put now on a firmer basis.   
The increased lattice dimensionality
that followed however allowed us to use only a limited 
number of lattice cites in each dimension, even though
larger lattices
would be better suited for a problem with such large scale
hierarchies as the present one, and they would possibly allow
for a better accuracy than $10\%$ in our solutions.

It is also true that the 
specific truncation of the $S-D$ equation that we chose,
as well as the dropping of the imaginary parts of all the
self-energies could influence these
 results non-trivially. Moreover,  the number of fermion flavours
 might not be large enough to justify the $1/N_{f}$ expansion
 and the one-loop diagrams we consider. We will try to 
 return to these issues in a future publication.

\noindent {\bf ACKNOWLEDGEMENTS} \\
I am indebted to I.J.R. Aitchison and M. Lindner for going through
the original draft of the paper and pointing out 
several issues. I also thank V. Miransky 
for helpful discussions. 
This research is supported by an {\it Alexander von Humboldt Fellowship}.

\end{document}